# Charged Three-Dimensional Solitons in a Ferroelectric Nematic Liquid Crystal

Netra Prasad Dhakal[1,2], Zakaria Siddiquee [1,3], Hiroya Nishikawa[4], Fumito Araoka[4], James Gleeson[3], Samuel Sprunt[1,3] and Antal Jákli[1,2,3]*

[1]Advanced Materials and Liquid Crystal Institute, Kent State University, Kent, OH 44242, USA

[2]Materials Science Graduate Program, Kent State University, Kent, OH 44242, USA

[3]Department of Physics, Kent State University, Kent, OH 44242, USA

[4] Physicochemical Soft Matter Research Team, RIKEN Center for Emergent Matter Science (CEMS), 2-1 Hirosawa, Wako, Saitama 351-0198, Japan

*: Corresponding author: ajakli@kent.edu

## Abstract

Experimental observations and computer simulation of formation and structural properties of so far experimentally never seen electrically charged propagating 3D solitons are reported. The propagating solitons appear spontaneously without an external field in a ferroelectric nematic ($N_F$) liquid crystal under rapid cooling from the antiferroelectric ($SmZ_A$) phase. The observed solitons propagate either parallel or antiparallel to the ferroelectric polarization and annihilate when they collide with each other. They are attributed to either positively or negatively charged propagating local polarization splay formed to screen excess negative or positive mobile charge lumps in the $N_F$ phase. The presence of mobile charge is attributed to a spatially periodic charge distribution present in the antiferroelectric phase that temporarily remains after the temperature is quenched through the antiferroelectric- ferroelectric phase transition.

## 1. Introduction

A "soliton", originally named by Zabusky and Kruskal (*1*), is a localized, finite-energy, nonlinear wave or field configuration that propagates without changing its shape. After the discovery of a one-dimensional soliton in water waves by Russell (*2*), solitons of various types have been observed ranging from biological systems (*3*) to cosmic strings (*4*). However, in three spatial dimensions there is a significant obstruction to "true" solitons that maintain their shape while propagating. The Derrick–Hobart theory (*5*, *6*) shows that for a large class of nonlinear field theories, a static, finite-energy, shape-preserving lump in 3D is generically unstable — it either collapses or spreads out — unless some other mechanism stabilizes it. Polar skyrmions, merons, torons and Hopfions have been observed in ferroelectric crystals (*7–9*) and in liquid crystals (*10–12*) but they are not propagating and their formation requires an external electric field.

Solitons observed in liquid crystals are typically one- and two-dimensional (*13*, *14*) and the recent examples of three-dimensional solitons in nematic liquid crystals are exclusively stabilized and driven by external electric fields (*15–19*). The recent demonstration of ferroelectricity in nematic fluids (*20*) comprising molecules with permanent dipole moments of ~10 D (*21–23*) opens a new avenue for investigating soliton phenomena. Ferroelectric nematic ($N_F$) liquid crystals possess a spontaneous polarization parallel to the axis of the nematic orientational order (director) and with magnitude as large as 8 μC/cm$^2$ (*24*), comparable to the polarization of ferroelectric crystals. These materials exhibit extraordinary physical properties such as ultra-low ($E \leq 1\,V/mm$) threshold for switching, large apparent dielectric constant, strong electrowetting, electromechanical instability, thermomechanical coupling, and spontaneous fluid filament formation (*25–34*). Non-propagating polar meron-like electric monopole (not a soliton) was observed in a chiral $N_F$ material (*35*) and polarization reorientation solitons were induced by in-plane electric fields in an $N_F$ material (*36*).

While the existence of charged solitons has been predicted by three-dimensional CS+BI Abelian gauge theory (*37*), to the best of our knowledge, there is to date no experimental evidence of charged, "true" 3D solitons in any field of science.

In this paper we present the first observation of shape preserving three-dimensional propagating solitons that form spontaneously in the well-studied $N_F$ material DIO (*22*) upon cooling from a higher temperature antiferroelectric state (designated alternatively as $SmZ_A$ (*38–40*) or splay nematic ($N_S$) (*41–43*)) to the $N_F$ phase. These solitons can be classified as charged (3+1)D solitons where 3 refers to the 3 dimensionality and 1 means they propagate in one dimension (*44*). Unlike the electrically-driven director bullets observed in nematic liquid crystals (*15–18*), they do not require external electric field to propagate, and their motion is either parallel or antiparallel to the ferroelectric polarization. Based on measurements on DIO doped with ionic liquids and on the results of simulations, we present evidence that these solitons form due to mobile ionic lumps screened by bound charges of localized divergence of the ferroelectric polarization field.

## 2. Experimental results

We studied pure 2,3′,4′,5′-tetrafluoro[1,1′-biphenyl]-4-yl 2,6-difluoro-4-(5-propyl-1,3-dioxan-2-yl) benzoate (DIO)(*22*) films between 1.7 $\mu m$ and 8.2 $\mu m$ thickness with boundary conditions such that the nematic director is uniform and parallel to the substrates ("planar" alignment). The molecular structure and temperature dependence of the spontaneous polarization of DIO are shown in Figure S1 of the Supporting Information. The material was first cooled at the rate of 1 $°C/min$, from 120 $°C$ in the nematic phase to 68 $°C$ in the antiferroelectric $SmZ_A$ phase that exhibits a weak density modulation perpendicular to the director. The films were then cooled at various rates between 0.2 $°C/min$ and 10 $°C/min$ to the $N_F$ phase. We observed the textures of the samples between crossed polarizer and analyzer by a polarizing optical microscope (POM).

Representative POM textures containing solitons in a $2.0 \pm 0.1$ $\mu m$ DIO film are shown in Figure 1. Figure 1(a) shows the texture at the $SmZ_A$-$N_F$ transition at $66.5°C$ while cooling the sample at 10 $°C/min$ rate. Except for a few inhomogeneities, the texture is dark indicating the director parallel to the polarizer. At about 0.1 °C below we observed the formation of several bright, elongated V-shaped domains, each a few microns in length and having similar appearance to the director bullets found in nematic films under AC electric fields (*15–18*).

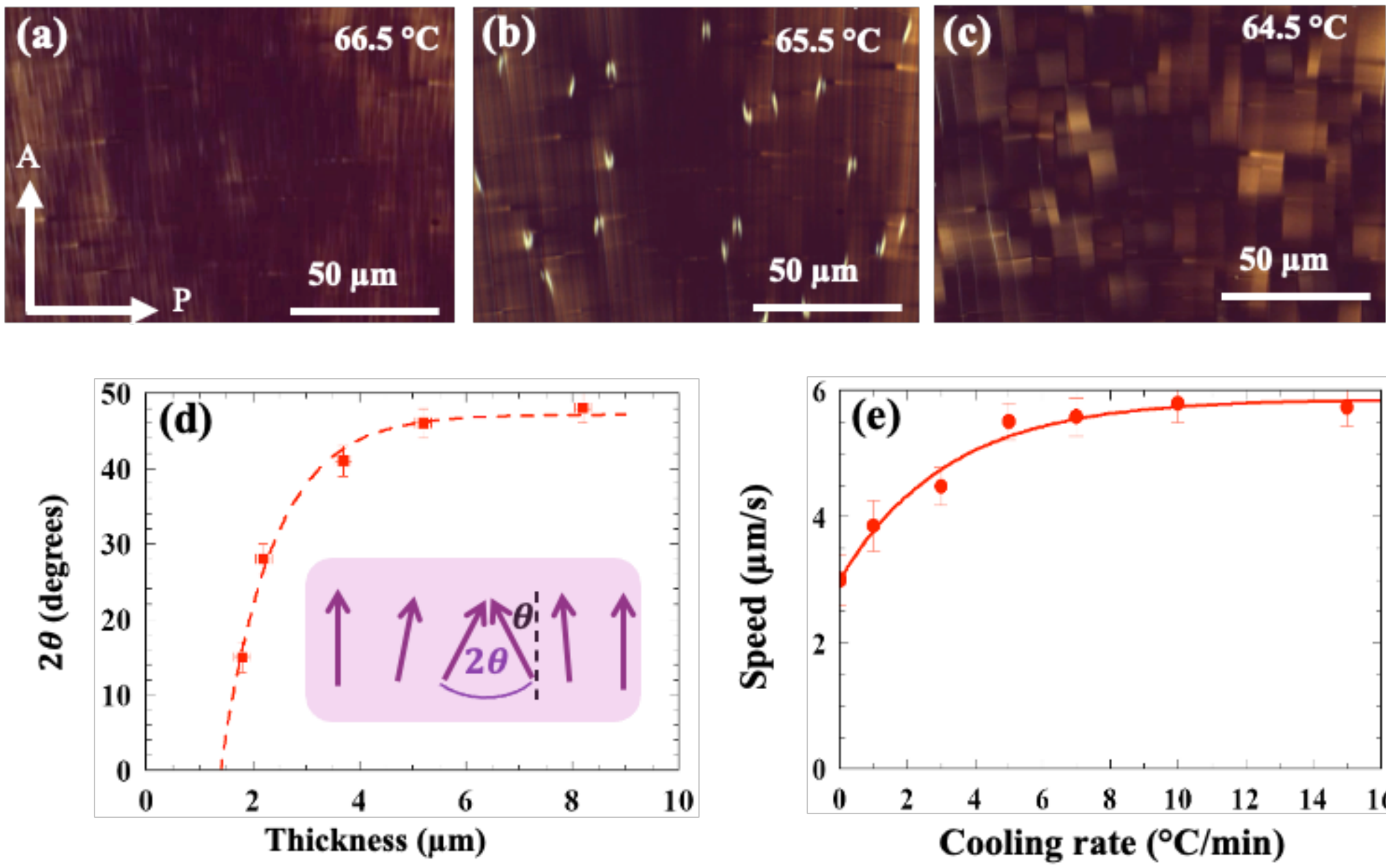

*Figure 1: Representative POM textures near solitons of a* $2.0 \pm 0.1$ µm *DIO film between crossed polarizers with the analyzer (A) parallel to the rubbing direction. (a) Texture at the* $SmZ_A$*-*$N_F$ *transition at 66.5°C while cooling the sample by 10 °C/min rate; (b) A snapsot of the texture at 65.5°C with several solitons; (c) Texture at 64.5°C with no solitons remaining; (d) Splay angle* $(2\theta)$ *as a function of film thickness; (e) Average speed of the solitons as function of cooling rate.*

As can be seen, these structures have a pointed, "Vee" shape. They are always observed to propagate parallel to the rubbing direction, but opposite to the direction of the Vee. We observe a roughly equal number of each Vee orientation. The formation and movement of these objects are shown in Supplementary Video 1. Although the shape of the solitons is similar to the "director bullets" previously observed in apolar nematic (N) liquid crystals, there are two important differences. First in the N phase the soliton moves perpendicular to the director, while the solitons we observed in the $N_F$ phase move along the director. Secondly, the director bullets in the N phase require external electric field to exist and propagate, whereas the solitons in the $N_F$ phase exist and move without an external field. Example images of these structures at 65.5 $°C$ are shown in Figure 1(b). Figure 1(c) shows a blocky texture at 64.5°C after all solitons have disappeared. This texture is much less uniform than that in Figure 1(a), indicating that the motion and collisions among oppositely charged solitons disturbs the local director orientation. Figure 1(d) shows the thickness dependence of the average angle $2\theta$ between the arms of the Vees. The director alignment angle $\theta$ in the arms has been determined by rotating the analyzer with respect to the crossed position until an extinction was seen. The definition of $2\theta$ is illustrated in the inset to Figure 1(d). The best fit of the measured thickness dependence of $2\theta$ is shown by dotted red line indicates a threshold thickness of about 1.5 μm and the saturation of the tilt angle of $2\theta \approx 45°$ above 5 μm thickness. Indeed textures of a $1.7 \pm 0.1\ \mu m$ film (see Figure S3(a) shows only very few and faint solitons. The thickness dependence of the shapes, brightness, and number density of the solitons shown in Figure S3(a-d) confirm this dependence. Figure 1(e) shows the effect of the cooling rate on the speed of the solitons for a $2.0 \pm 0.1\ \mu m$ thick DIO film. The speed increases from $\sim 3\ \mu m/s$ at constant temperature up to $\sim 6\ \mu m/s$ under 10 °C/min cooling rate. The speed was calculated from the measure the distance the solitons travel in a fixed time so that the end of the measurement the temperature was $\sim 65.0\ °C$. Therefore, for higher cooling rates the starting and average temperature is higher while the average viscosity is lower.

Figure 2 (a-c) display higher magnification images of three different optical configurations. Figure 2(a) corresponds to the rubbing direction at 45° to the polarizer and analyzer. Figure 2(b) is the same but with a wave plate inserted, and Figure 2(c) also with wave plate but the rubbing direction is parallel to the analyzer. Comparing all three of these, and mapping the interference colors to the Michel-Levy chart and relying upon the published birefringence values for DIO (*45*), we can conclude the following (see details in Section III of the Supporting Information): (i) the director in the Vee structures is rotated out of the substrate plane; (ii) the opposing Vee arms have equal but opposite director rotation away from the rubbing direction of about 12°. This director configuration represents a splay deformation. However, in the $N_F$ phase splay is highly disfavored as it results in a highly energetic non-zero charge density, $\rho = -\vec{\nabla} \cdot \vec{P}$. The implications of this are discussed in the Discussion.

We also observed interactions between two solitons moving in opposite directions. To illustrate this, in Figure 2(d and e) we show enlarged snapshots taken in ~10s intervals of two Vees approaching each other. Figure 2(d) shows an interactions as the right arm of the upper Vee (moving downward) encounters the left arm of the lower Vee (moving upward). In this example the left and right arms of the oppositely moving Vees momentarily form an "N" shape, which then evolves into a stationary, uniform texture. That is, the two counterpropagating Vees annihilate each other. Figure S2(a-c) of the supporting information show the spatial dependence of the transmitted light intensity of the Vees $10s$ before (Figure S2(a), during (Figure S2(b)) and $10s$ after the collison (Figure S2(c)). One can see how the solitons merge, and then the intensity profile drops to uniform. Figure 2(e) shows snapsots taken in $10s$ intervals to illustrate a collision of a pair of Vees that travel toward each other along the same center axis. In this case the solitons fade away and disappear as the distance between them becomes less than their $\sim 5\ \mu m$ lengths. These are consistent with the scenario that opposing Vees are solitons having nearly equal and opposite charges that become screened out when the Vees meet. This framework is in contrast to neutral solitons that typically pass through each other.

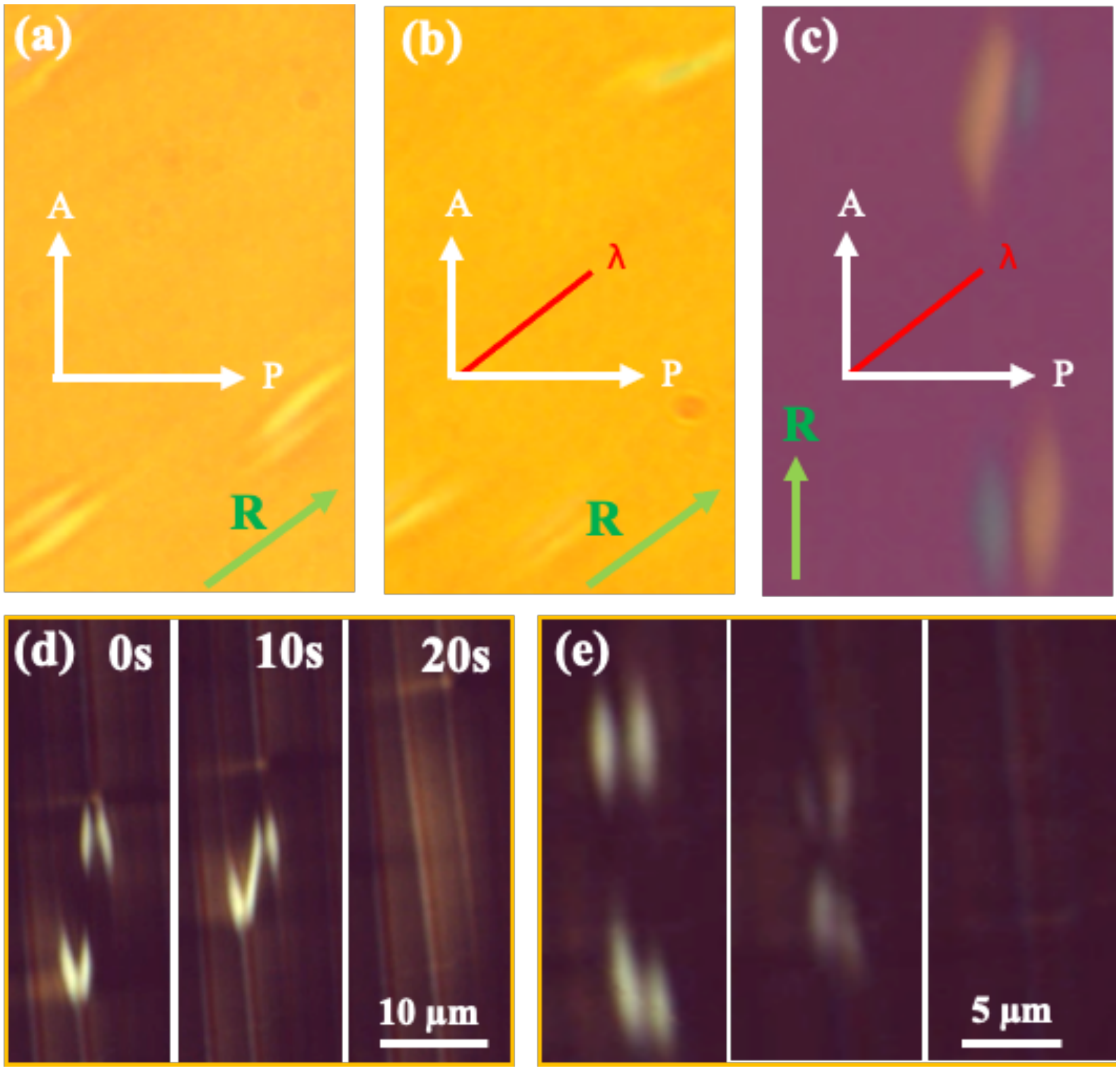

*Figure 2: Detailed structures of the Vee shaped solitons. (a,b) Textures at 65.5°C between crossed polarizers with director 45° away from the polarizer showing solitons without (a) and with a λ-plate (b), and when a λ-plate is inserted at 45° with respect to the polarizer (e); Textures at 65.5°C between crossed polarizers when the analyzer is parallel to the director and when a λ-plate is inserted at 45° with respect to the polarizer. (d) Sequential images of counter-propagating not collinear soliton pairs to illustrate their interaction; (e) Sequential of collinear counter-propagating interacting solitons.*

Figure S4 demonstrates the effect of added free ions on the soliton structure. The antiferroelectric $SmZ_A$-$N_F$ phase transition temperature and the angle of the director between the soliton arms are shown functions of the concentration of a miscible ionic liquid (EMIM-TFSI) in Figure S4(a). Representative textures at 0.016% and 0.08% EMIM-TFSI concentrations are shown in Figure S4(b,c) and Figure S4(d), respectively. At 0.016% ionic liquid concentration, the number of solitons are higher than in the pure DIO and the oppositely moving solitons do not annihilate but they seize together on close approach and form biconvex lens shaped structures (Figure S4(c) and Supplementary Video 2). In contrast to the pure DIO and the 0.016 % EMIM-TFSI doped DIO sample, the 0.08% EMIM-TFSI doped DIO sample contains only non-equilibrium focal-conic defects and no solitons (see Figure S4(d)).

Finally, we have searched for the appearance of propagating charged solitons in an another well-known $N_F$ material, RM734 (*23*) which has no (or only a very narrow) antiferroelectric range. We have also tested a room temperature $N_F$ mixture FNLC 919 from Merck with an antiferroelectric range above the $N_F$ phase comparable to that of DIO.(*40*) No solitons were observed on cooling to $N_F$ phase in either of these materials.

## 3. Discussion

The observations that the oppositely moving solitons annihilate each other and they have opposite director splay, i.e., opposite sign bound charge densities, support the idea that the solitons are charged and the oppositely moving solitons have different charges. However, nonzero charge density imposes an energy cost proportional to $P^2$. Thus, one expects locally charged structures to be strongly suppressed unless a compensating free charge is present. We propose this is possible owing to the remnant distribution of mobile ionic species after rapid cooling through transition from the antiferroelectric $SmZ_A$ to the $N_F$. Specifically, the nanoscale alternation of $\boldsymbol{P}$ in the antiferroelectric phase can induce a modulation of excess ionic densities due to periodic splay (*41–43*) and/or electron density variation (*38–40*). At the transition to the $N_F$ phase in pure DIO, the periodicity of the nanoscale modulations becomes infinite, but the ions can only attain their ultimate equilibrium via diffusion. If the phase transition does not proceed quasi-statically, one expects regions of either excess free charge. This is the charge which we propose is compensated with local, induced bound charges associated with non-zero splay.

This model is supported by the observation that below a threshold thickness (~1.5 μm, Figure 1(d)) there are no solitons, since in thin films the time for neutralization is shorter than the formation of the ferroelectric structure. It also explains why at very low (0.016%) concentrations of ionic liquid increased the number of the solitons, while at higher (0.08%) ionic liquid concentration there are no solitons. At low excess ion concentrations the $SmZ_A$-$N_F$ transition is still sharp enough to complete the transition before reaching a neutral ion distribution with an increased number of excess ions; whereas at high ion concentration, the transition is not sharp anymore so the ions could reach the neutral state before the transition to the $N_F$ phase is completed.

Finally, it also explains the lack of solitons in a different ferroelectric material (RM734) that does not exhibit a wide antiferroelectric phase, and in FNLC919 with wide $SmZ_A$ phase(*40*) but with very different temperature dependences of the wavenumber $q_s$ of the density modulation than in DIO (*46*). Within the antiferroelectric range, the wavenumber in DIO only barely changes from $q_s \approx 0.72\, nm^{-1}$ to $q_s \approx 0.68\, nm^{-1}$ (corresponding to a small increase in layer spacing from $L_s = 8.7\, nm$ to $L_s = 9.2\, nm$) and then abruptly drops to zero at the transition to the $N_F$ phase. On the other hand, in the FNLC 919 the wavenumber varies considerably from $q_s \approx 0.82\, nm^{-1}$ to $q_s \approx 0.38\, nm^{-1}$ (corresponding to an increase in $L_s$ from $7.7\, nm$ to $16.5\, nm$) within the lower 3 °C range of the $SmZ_A$ phase, indicating a more gradual and possibly continuous divergence of the layer spacing on approaching the transition to the $N_F$ phase. Additionally, the phase transition is much sharper in the single compound DIO than on the FNLC919 mixture thus allowing much less time for the mobile charges to neutralize. This mechanism also explains why no soliton was found in the 0.08% EMIM-TFSI doped DIO sample, where the phase transition is not sharp anymore.

Based on this physical mechanism and on the textures shown in Figure 1 and Figure 2, a proposed soliton composition is sketched in Figure 3(a). As deduced from Figure 2(a), the director in the arms of the solitons tilt opposite to the tilt of the arms. Assuming the rubbing direction is along the $y$ axis (up), the polarization is pointing to the opposite direction for DIO (*47*). For positive excess mobile charge, the splay is producing negative charge, as illustrated on the left side of Figure 3(a). When the arms of the Vees are pointing up, i.e., the divergence of the downward pointing polarization is decreasing upward, the solitons move opposite to $\boldsymbol{P}$ (along the rubbing direction). For negative remnant mobile charge, all directions are reversed and the solitons move in the direction of $\boldsymbol{P}$.

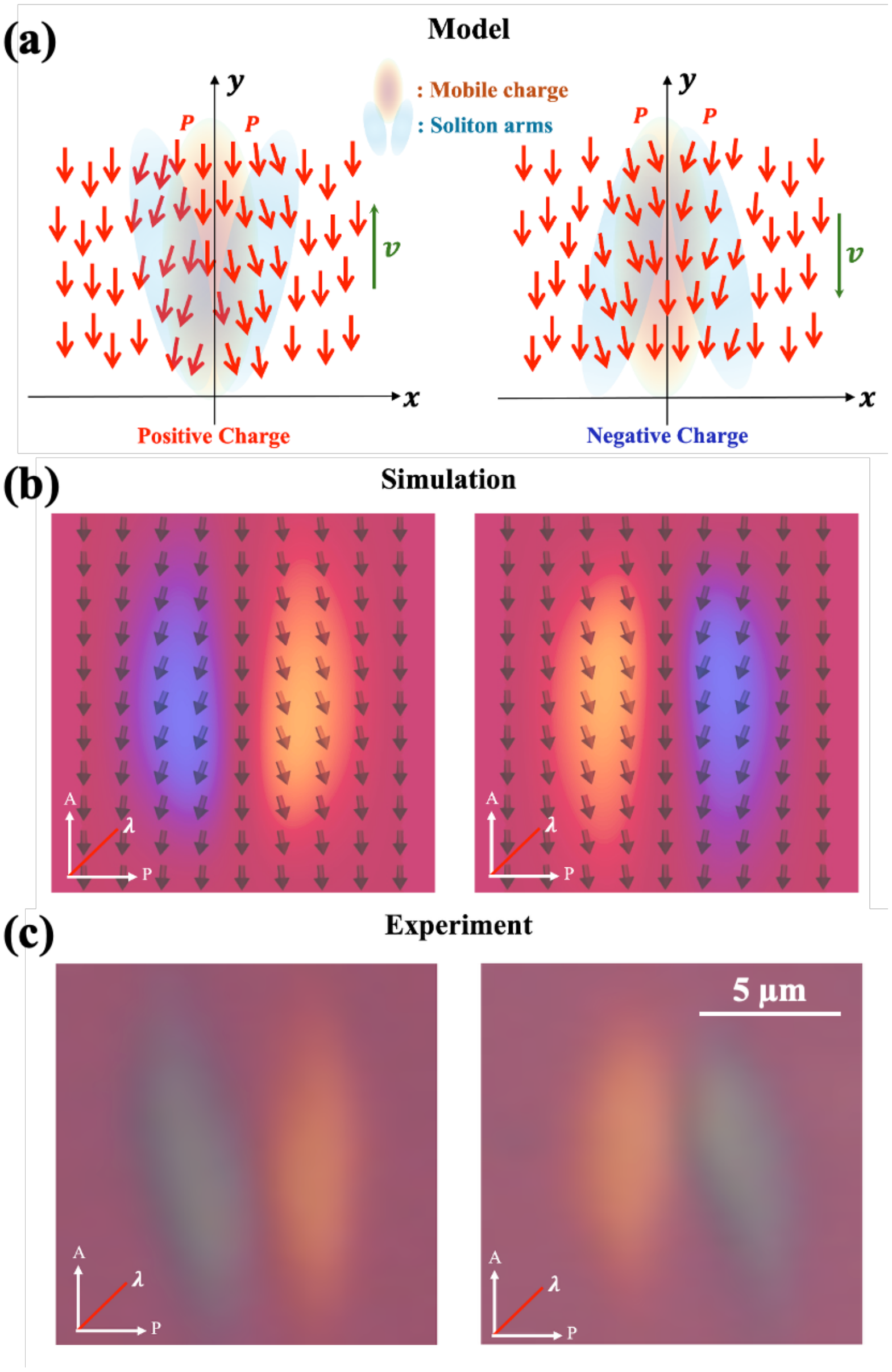

*Figure 3: Illustration of the polarization structure of solitons forming around positive (left) and negative (right) excess charges. (a) A qualitative model based on polarized optical microscopy (POM) observations and the physical mechanism discussed above; (b) Simulated POM textures of a soliton with overlayed polarization structure when a λ plate is inserted at 45° with respect to the crossed polarizers; (c) Actual POM textures observed in experiments with the same conditions as in (b).*

We have simulated the director, the polarization and the electrostatics as described in sections VI-VIII of the Supporting Information. This checks whether the above physical mechanism is a reasonable description of the experimental results. From the numerical solutions we derive simulated POM textures and polarization fields as shown in Figure 3(b). Elongated arms with size, shape and birefringence colors strongly resemble those found experimentally (Figure 3(c)). As shown in Supplementary Video 3, the simulated movement of the solitons via propagation of the polarization splay accurately describes the direction and speed of the experimentally recorded motion when the splay elastic constant is 2 orders of magnitude larger than of the twist or bend elastic constants. Such large value is due to the splay induced bound charge that increase the energy cost of a splay, thus increases the effective splay constant. (*20*, *47–49*)

The simulation also shows that the slight V-shape of the soliton arms is due to a non-zero, splay flexoelectric coefficient, $\gamma_f = e_1 \varepsilon_o \varepsilon$, where $e_1$ is the splay flexoelectric constant as defined by Meyer (*50*) and $\varepsilon_o \varepsilon$ is the dielectric permittivity. $\gamma_f < 0$ lowers the energy of a polarization converging along its own direction, whereas $\gamma_f > 0$ favors a diverging polarization. The best agreement with the experimentally observed shape was obtained with $\gamma_f = -0.006\ V$. This value is comparable to that used by Rupnik et al (*51*). These agreements between the simulations and the experimental results support of our proposed physical mechanism about the formation and structure of the charged solitons.

## 4. Conclusions

To summarize, we observed the formation of electrically charged, propagating solitons in ferroelectric nematic liquid crystals after rapid cooling from the antiferroelectric phase. Solitons carrying opposite bound charges and move in opposite directions and annihilate when they collide. These charged 3D solitons exist and propagate due to the combination of ferroelectric polarization and 3D fluidity. As verified by our computer simulation, solitons with polarization splay form to screen out excess mobile charges that appear when the mobile charges of the antiferroelectric phase do not have time to neutralize during a sharp antiferroelectric – ferroelectric transition. We conjecture that such conditions occur only in materials with low density of impurity ions such as DIO where the transition is first order.

## 5. Acknowledgments

This work was supported by the US National Science Foundation grant no. DMR-2210083.

# Supporting Information

# Charged Three-Dimensional Solitons in a Ferroelectric Nematic Liquid Crystal

Netra Prasad Dhakal[1,2], Zakaria Siddiquee [1,3], Hiroya Nishikawa[4], Fumito Araoka[4], James Gleeson[3], Samuel Sprunt[1,3] and Antal Jákli[1,2,3]*

[1]Advanced Materials and Liquid Crystal Institute, Kent State University, Kent, OH 44242, USA

[2]Materials Science Graduate Program, Kent State University, Kent, OH 44242, USA

[3]Department of Physics, Kent State University, Kent, OH 44242, USA

[4] Physicochemical Soft Matter Research Team, RIKEN Center for Emergent Matter Science (CEMS), 2-1 Hirosawa, Wako, Saitama 351-0198, Japan

*: Corresponding author: ajakli@kent.edu

## I. Materials and Methods

In our experiments we used the standard ferroelectric nematic compound 2,3′,4′,5′-tetrafluoro[1,1′-biphenyl]-4-yl 2,6-difluoro-4-(5-propyl-1,3-dioxan-2-yl) benzoate (DIO) (*1*). This material's phase sequence on cooling is I 158 °C N 84.5 °C $SmZ_A$ 68.8 °C $N_F$. $SmZ_A$ is an antiferroelectric phase (*2*), also called as splay nematic ($N_s$) phase (*3*) . The molecular structure and the temperature dependence of the ferroelectric polarization of DIO are shown in Figure S1. For some measurements we added an ionic liquid to DIO: the molecular structure of the ionic liquid chosen is shown in Figure S4(a).

All materials were filled into planar cells with gaps between 1.7 μm and 8.2 μm at 120 °C by capillary action along the rubbing direction. The cells were constructed by assembling two glass substrates coated with PI2555 and unidirectionally rubbed, which gives uniform planar alignment with director in the rubbing direction.

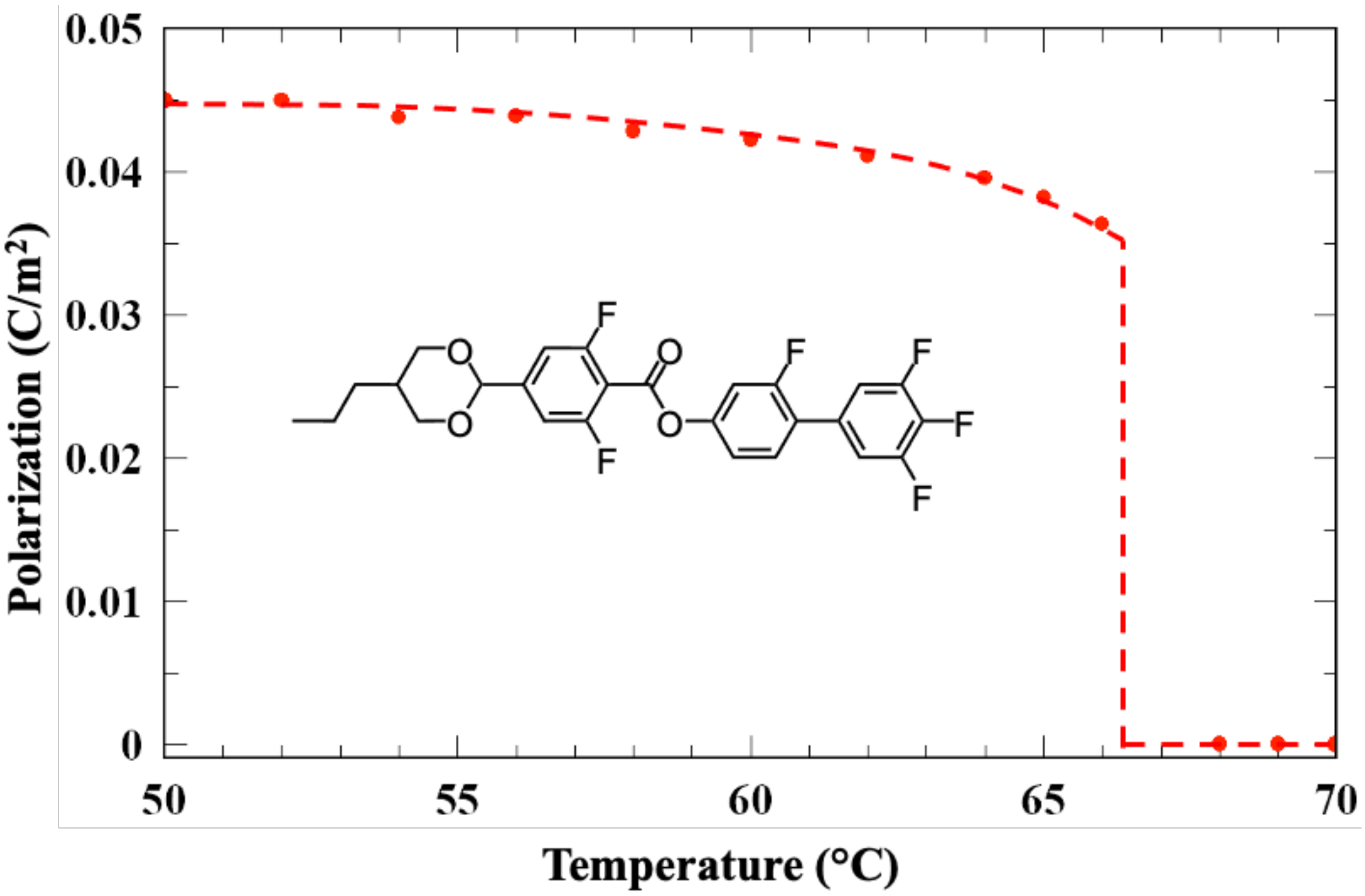

*Figure S1: Molecular structure and the temperature dependence of the ferroelectric polarization in the $N_F$ phase of pure DIO. The saturated polarization is not measured in the antiferroelectric phase.*

## II. Annihilation of solitons with opposite charges

Spatial dependence of the transmitted light intensity showing the annihilation of colliding solitons before during and after interaction is shown in Figure S1.

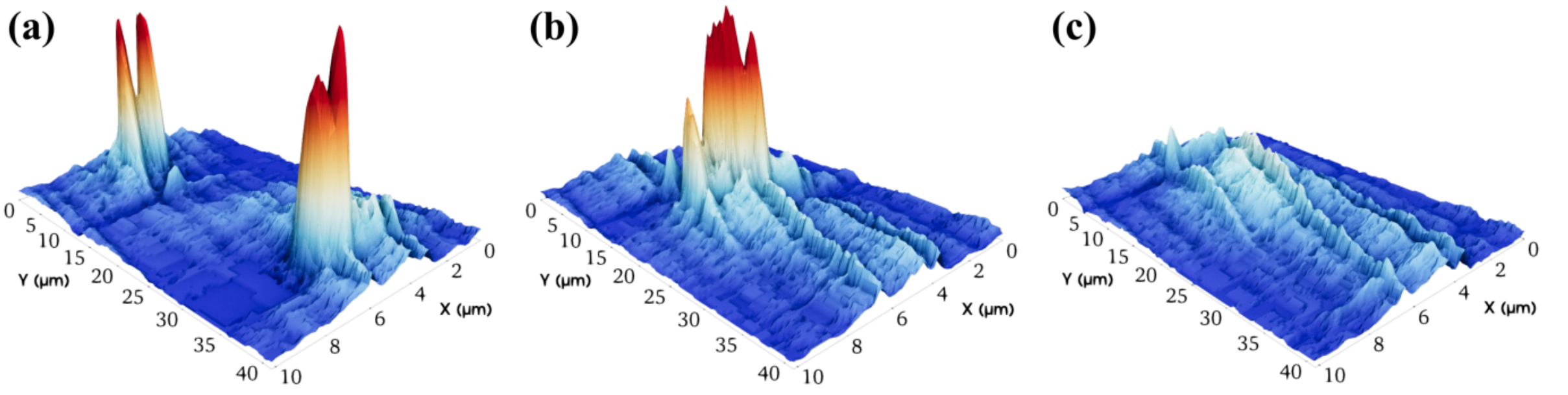

*Figure S2: Spatial dependence of the transmitted light intensity at the solitons* 10*s before (a), during (b) and* 10*s after(c) the interaction.*

## III. Analysis of the texture with $\lambda$ plate

As seen in Figure 2(c), the texture outside the solitons has rich maroon color corresponding to $\Gamma \sim 640 \pm 10\ nm$ optical path difference. Considering that total path difference of two retarders with fast axes making an angle $\theta$ with each other is

$$\Gamma = \sqrt{\Gamma_\lambda^2 + \Gamma_{LC}^2 + 2\Gamma_\lambda \Gamma_{LC} cos2\theta}, \qquad (S1)$$

and for the area outside the solitons $\theta \approx 45°$, we get from (S1) that $\Gamma_{LC} \approx \sqrt{\Gamma^2 - \Gamma_\lambda^2} \approx \sqrt{(630 \pm 10)^2 - 530^2} \approx 360 \pm 20\ nm$. Within the uncertainties of the measurements, this value agrees with $\Gamma_{LC} \approx 380\ nm$ deduced from Figure 2(a). This picture also shows that the left and right arms of the solitons with respect to their movement direction have different colors: medium navy and tangerine colors, respectively. This indicates that the director in the different arms of the solitons are oppositely tilted with respect to the lambda plate and the optic axis of the director outside the solitons. The navy and tangerine colors roughly correspond to $\Gamma \sim 720\ nm\ (460\ nm)$. From (S1) we get that $\Gamma \sim 720\ nm$ i.e., $cos2\theta|_{navy} \sim \frac{530^2+350^2-720^2}{2 \cdot 530 \cdot 350} \approx -0.31$ wich corresponds $\theta(navy) \sim 54°$. Similarly, $\Gamma \sim 460\ nm$ means $cos2\theta|_{tangerine} \sim \frac{530^2+350^2-460^2}{2 \cdot 530 \cdot 350} \sim 0.52$ wich corresponds $\theta(tangerine) \approx 30°$. This means that the angle between the arms is $2\theta \approx 24°$ and the right and left arms tilt by $\theta \sim \pm 12°$ with respect to the uniform director (rubbing direction). This is in addition to their tilt with respect to the substrates. We note that the measured $|\Delta\theta|$ of the director is close to the average angle of the arm with respect to the rubbing direction (see Figure 1(d)) but it has the opposite sign, i.e., $\theta_n \approx -\theta_{arm}$.

## IV. Effect of film thickness

Typical textures with solitons are shown in Figure S3. Textures of a $1.7 \pm 0.1\ \mu m$ film (see Figure S3(a) shows only very few and faint solitons. The brightness and the width of the solitons increase with increasing thickness as seen in textures of $2.0 \pm 0.1\ \mu m$ (Figure S3(b)), $3.7 \pm 0.1\ \mu m$ (Figure S3(c)) and $5.2 \pm 0.1\ \mu m$ (Figure S3(d)). It is also seen that larger the film thickness, the larger the effect of the solitons on the overall alignment, i.e., the traces showing the previous movement of the arms of the solitons are more pronounced. These are most noticeable for the 3.7 μm and 5.2 μm thick films and likely related to the larger splay deformation inside the solitons moving along the rubbing direction.

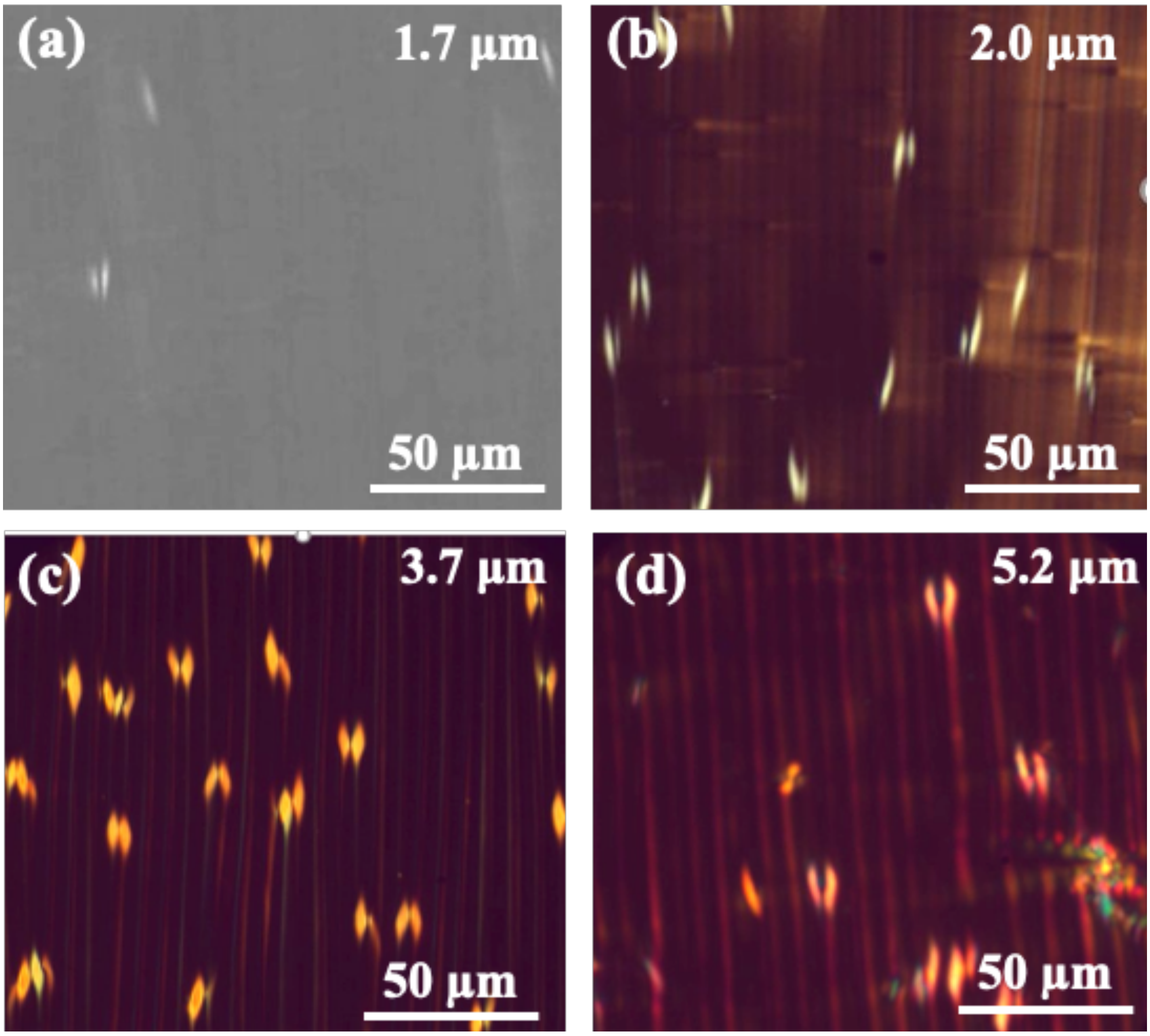

*Figure S3: Effect of film thickness on the structure at 65.5 °C. (a) Texture of an* $1.7 \pm 0.1$ μm *film;(b)Texture of a* $2.0 \pm 0.1$ μm *film; (c) Texture of a* $3.7 \pm 0.1$ μm*film; (d) Texture of a* $5.2 \pm 0.1$ μm *film.*

## V. Effect of Additional Ions

As seen in Figure S4(a), the $SmZ_A$-$N_F$ phase transition temperature linearly decreases with the concentration of EMIM-TFSI from 66°C at 0% to 57°C at 0.08 % ionic liquid concentration. At the same time, the angle $2\theta$ between the arms at the tip of the solitons for (averaged over 10 solitons for the Pure DIO and the 0.016%EMI-TFSI doped DIO) samples and of the focal conics for the 0.08% mixture, increases roughly linearly with the ion concentration. While for the 2 μm thick pure DIO sample $2\theta \approx 24 \pm 3°$ (see Figure 1(d)), for the 0.016 % and 0.08 % ion doped samples $2\theta \approx 39 \pm 3°$ and $2\theta \approx 80 \pm 3°$ were found, respectively.

POM textures formed at 0.016% and 0.08 % ionic liquid concentrations for 2.0 μm films are shown in Figure S4(b-d). By the addition of 0.016% ions (Figure S4(b)), the number density of solitons increases by about 50% compared to the pure DIO sample. The formation and propagation of the solitons is also shown in the Supplementary Video 2. Just as seen for thicker pure DIO samples, the presence and motion of the solitons also leads to a stripe texture with defect walls ending on the solitons. However, while for the pure DIO the solitons gradually disappear, in the 0.016% EMIM-TFSI samples the oppositely moving solitons cease as they get close to each other, and form biconvex lens shape structures as shown in Figure S4(c).

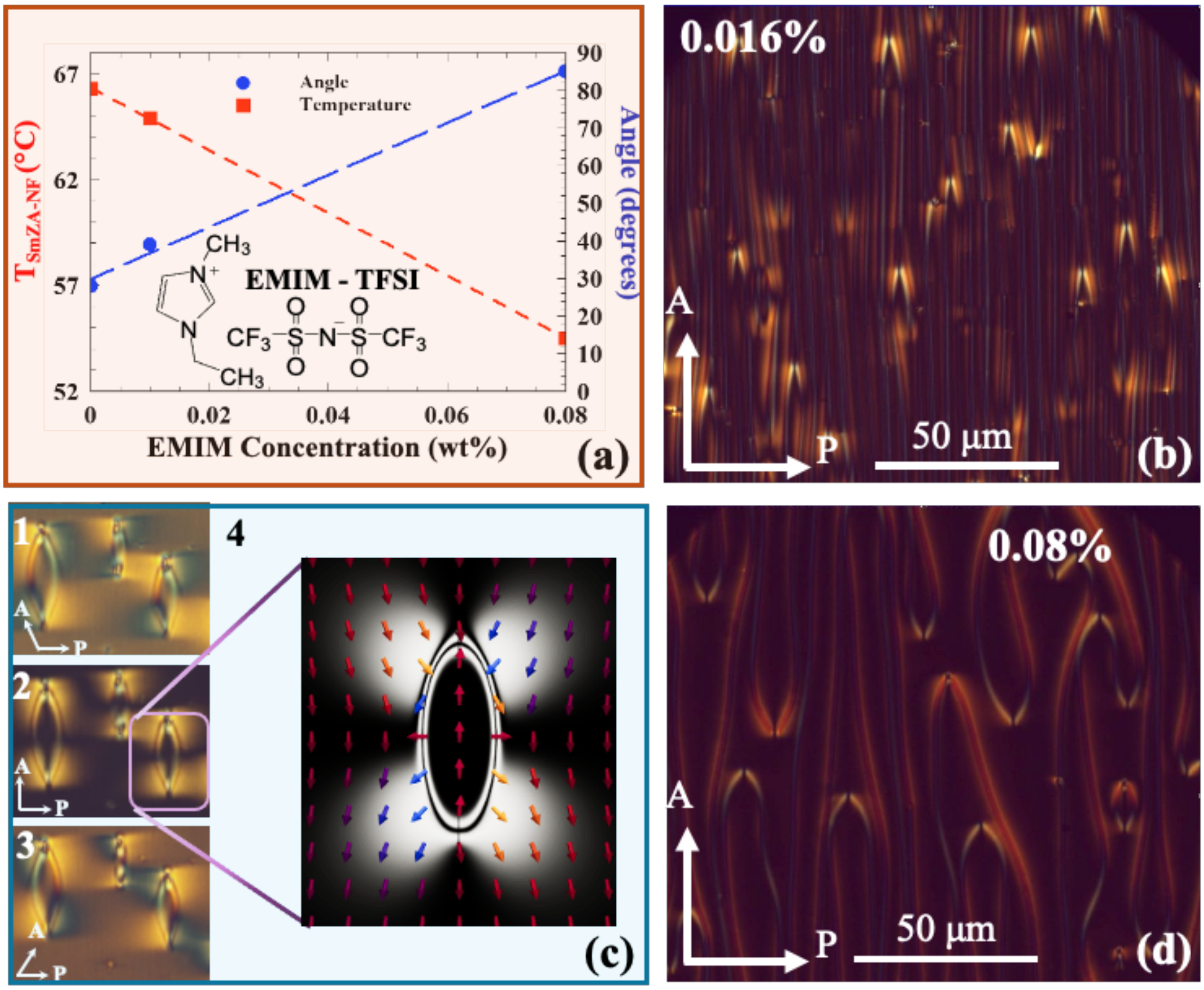

*Figure S4: Summary of the effect of the ionic liquid EMIM-TFSI on the structures of solitons. (a) $SmZ_A$-$N_F$ phase transition temeprature (left axis) and director angle between the different arms of the solitons (right angle) as function of th ionic liquid concentration in DIO; (b) POM texture with solitons for 0.016 % EMIM-TSFI doped DIO; (c) POM textures of still soliton pairs for 0.016 % EMIM-TFSI doped DIO with counterclockwise uncrossed (1), crossed (2) and clockwise uncrossed (3) analyzer, and (4) with a sketch of the simulated polar dircetor structure near the lens-shaped soliton pair; (d) POM texture for a 0.08 % EMIM-TFSI doped DIO.*

Selected pairs of seazed solitons are seen in Figure S4(c1-c3) between crossed polarizers (Figure S4(c2) and Supplymentary Video 2). Under crossed polarizers, the texture is dark far away and inside the lenses indicating director parallel to the analyzer. Right above and below the lenses, the texture is bright indicating director tilted away from the vertical analyzer. Looking at the texture when the analyzer is rotated by $20°$ counterclockwise (CCW) in Figure S4(c1) and clockwise (CW) in Figure S4(c3), we see that the upper-right and bottom-left parts outside the lens become dark under CCW rotation, while the upper-left and bottom-right parts become dark under CW rotation. The dark regions indicate areas where the director is perpendicular to the analyzer. Such tilt directions are sketched in Figure S4(c4) and are consistent with that deduced from observations with $\lambda$ – plate near the solitons of the pure DIO (see Figure 1(c)). In addition to the director structure, in Figure S4(c4) we also included the polarization directions and defects.

Representative POM texture of a 0.08 % EMIM-TFSI doped DIO sample is seen in Figure S4(d). One can see no solitons but only focal-conic type domains. The angle between the sides of the tip of the focal-conics becomes close to 90°, i.e., the bend and splay have equal contributions. In contrast to the previous cases when the V-shaped solitons moved along the diverging arms, now they are moving toward the converging ends of the arms. Also, while previously oppositely moving solitons annihilated as they were colliding, now they avoid each other by squeezing between two oppositely moving domains. Finally, after the focal conics find their equilibrium positions, the texture becomes uniform, instead of leaving traces of the previous solitons. This confirms that, in contrast to the pure DIO and the 0.016 % EMIM-TFSI doped DIO sample, the 0.08% EMIM-TFSI doped DIO sample contains only non-equlibrium focal-conic defects while finding their equilibrium uniform alignment.

## VI. Simulation Model

For the free energy density, we use the expression developed for achiral ferroelectric nematic materials by Zou and Aya (19), without its Landau bulk free energy term. In the non-polar nematic phase, the natural field variable is the headless unit vector, the director $\hat{n}$. In the ferroelectric nematic phase, the unit vector is polar, $\hat{n} = \boldsymbol{P} / |\boldsymbol{P}|$, and the magnitude of the polarization is the saturation polarization (19). We carry $\boldsymbol{P}$ itself as the unknown in units of $C/m^2$, and recover the director from it wherever an orientation is needed. With $K_1$, $K_2$ and $K_3$ the splay, twist and bend Frank constants (8, 9), the elastic component of the free energy density is:

$$f_{\mathrm{F}} = \frac{1}{2}K_1(\nabla\cdot\mathbf{n})^2 + \frac{1}{2}K_2(\mathbf{n}\cdot(\nabla\times\mathbf{n}))^2 + \frac{1}{2}K_3(\mathbf{n}\times(\nabla\times\mathbf{n}))^2 \quad \text{(S1)}$$

In the ferroelectric nematic phase, the splay elastic constant is the largest due to the electrostatic contribution of the bound charge density $\rho = -\vec{\nabla}\cdot\vec{P}$ associated with the splay deformation of the polarization. It provides the screening since neutralizing a packet of free charge is precisely a demand for divergence in **P**. The competition between the electrostatic benefit of screening and the elastic cost of the splay sets the size and shape of the texture around a free charge packet. Screening a positive (negative) packet demands a radially outward (inward) $\boldsymbol{P}$.

Because the ferroelectric nematic phase lacks head-tail symmetry, the elastic free energy density acquires a term linear in splay, the flexoelectric coupling term (19)

$$f_{\mathrm{flexo}} = -\gamma_f(\mathbf{n}\cdot\mathbf{P})(\nabla\cdot\mathbf{n}) \quad \text{(S2)}$$

where $\gamma_f$ is the bare splay flexoelectric coefficient. Its sign selects which splay is favored relative to **P**: $\gamma_f < 0$ favors $\nabla\cdot\mathbf{P} < 0$, which carries positive bound charge, whereas $\gamma_f > 0$ favors $\nabla\cdot\mathbf{P} > 0$. No measurement of $\gamma_f$ has been reported for DIO. For RM734 $|\gamma_f| \approx 0.006$ V and -0.01$V$ were used in (20), and in (21). We use $\gamma_f = -0.004$ V, where the negative sign favors converging splay in agreement with our experiment observation.

Spatial variations of the polarization, in both magnitude and direction, are penalized by the polarization gradient term described in Ref. (19)

$$f_{\mathrm{G}} = \frac{1}{2}h|\nabla\mathbf{P}|^2 \quad \text{(S3)}$$

with $h = \frac{K_{\mathrm{h}}}{{P_{\mathrm{B}}}^2}$ and $K_{\mathrm{h}} = 50$ pN and $P_B \approx 0.038\ C/m^2$ bulk spontaneous polarization of DIO at 65°C. Using this value of $K_{\mathrm{h}}$ together with the surface anchoring and the virtual layers described below, $|\boldsymbol{P}|$remains close to $P_B$, so the Landau bulk term $\frac{1}{2}A|\mathbf{P}|^2 + \frac{1}{2}B|\mathbf{P}|^4$ of ref. (19) is not needed.

As the thickness of the sandwich cell is much smaller than its lateral size, we model the film as a two-dimensional sheet. The anchoring energy per unit area has an apolar term of strength $W_{\mathrm{S}}$ and a polar part of strength $W_{\mathrm{P}}$ (10, 19). After division by the sample thickness $L_{\mathrm{z}}$ it gives the energy density $f_{\mathrm{A}} = \frac{1}{L_{\mathrm{z}}}\left[\frac{1}{2}W_{\mathrm{S}}(1 - (\mathbf{R}\cdot\mathbf{n})^2) - W_{\mathrm{P}}(\mathbf{R}\cdot\mathbf{n} - 1)\right]$, where **R** is the unit vector along the rubbing direction. The apolar term is invariant under $\mathbf{n} \rightarrow -\mathbf{n}$ and only sets the easy axis $\pm\mathbf{R}$, while the polar term breaks the head-tail symmetry and selects the sense of **P** along it. $W_P > 0$ favors $\mathbf{n} = \mathbf{R}$ and $W_P < 0$ favors $\mathbf{n} = -\mathbf{R}$. Both terms act on the direction of **P** only, not on its magnitude. The polarization at the surface $\mathbf{P_S}$ for DIO is known to align antiparallel to the rubbing direction on a rubbed polyimide surface (11). With the rubbing direction ***R*** taken along the positive y axis of the simulation domain, $\mathbf{P_S} = -P_0\mathbf{R} = (0, -P_0, 0)$.

To describe the screening of free charge by a polarization splay that creates a bound charge density $\nabla\mathbf{P} = -\rho_B$, we use the anisotropic dielectric tensor $\varepsilon_{ij} = \varepsilon_0\,(\varepsilon_\perp\delta_{ij} + (\varepsilon_\parallel - \varepsilon_\perp)\,n_i n_j)$. Screening of a free charge is treated in the Debye-Hückel approximation (16-18), which in the limit of $eV << k_BT$ assumes $\nabla \cdot (\hat{\varepsilon}\nabla V) - \frac{\varepsilon_{iso}\varepsilon_0}{\lambda_D^2} V = -\rho_f$, where $\varepsilon_{Iso}$ is the orientationally averaged permittivity and $\lambda_D$ is the Debye screening length.

Collecting all electric contributions, the electric free energy density becomes

$$f_E = -\mathbf{E}\cdot\mathbf{P} - \frac{1}{2}\mathbf{E}\cdot(\hat{\varepsilon}\mathbf{P}\cdot\mathbf{E}) - \frac{1}{2}\frac{\varepsilon_{iso}\varepsilon_0}{\lambda_D^2}V^2 + \rho_f V, \qquad \text{(S4)}$$

where $\mathbf{E} = -\nabla V$. The term containing $\lambda_D$ is the linearized Debye-Hückel free energy density of a background population of mobile ions, which screen linearly and locally rather than by deforming the director. Its stationarity with respect to $V$ supplies the screening term in the Poisson equation below. A stationary solution of $f_E$ with respect to V gives the Poisson equation,

$$\nabla\cdot(\hat{\varepsilon}\mathbf{P}\cdot\nabla V) - \nabla\cdot P - \frac{\varepsilon_{iso}\varepsilon_0}{\lambda_D^2}V + \rho_b = 0 \qquad \text{(S5)}$$

The free charge is represented as a single rigid, conserved packet. It is an ellipse of total charge $Q_f$, taken uniform through the film thickness, with a profile that falls smoothly to zero at the edge of the ellipse $\rho_f = A(1-u^2)^2$ for $u < 1$ *and* $\rho_f = 0$ for $u \geq 1$, where $u^2 = (\frac{x-x_c}{a})^2 + (\frac{y-y_c}{b})^2$. The elongation along the rubbing axis with the semi-axis $b$ taken twice $a$ reflects the anisotropy of the medium rather than any measured shape. The profile is continuously differentiable at $u = 1$, so the fixed quadrature rule has no step or kink to integrate. The amplitude A is recomputed numerically on every configuration so that $L_z$ times the integral of $\rho_f$ over the sheet equals $Q_f$ exactly on the given grid, which makes the charge conserved to machine precision and independent of where the packet sits relative to the lattice. The semi-axes a and b are frozen, so the only coordinate the packet carries is its position $(x_c, y_c)$.

Screening the packet requires a polarization of magnitude $Q/2\pi L_z r$ at a distance r, so the shorter semi-axis must stay above the saturation radius $r_s = \frac{Q_f}{2\pi L_z P_B}$ or the charge cannot be screened at the edge of its own cloud, $|\mathbf{P}|$ runs past $P_B$, and the director around it scrambles.

Physically this means that only so much charge can be held in so small a volume before the polarization of the host can no longer neutralize it, and the texture that results is a scrambled core rather than a soliton.

The model is implemented in DOLFINx 0.10.0 (1-4), with PETSc (5, 6) providing the linear and nonlinear algebra backend and dolfinx_mpc 0.10.5 (7) supplying the periodic constraints. The liquid crystal evolves toward the configuration that minimizes its total free energy, analogous to a mechanical system settling into its lowest potential energy state. The scalar free energy function is the volume integral of the total free energy density $f$ that quantifies the elastic distortions, the flexoelectric coupling between splay and polarization, the cost of spatial gradients of the polarization, the surface anchoring, and the electrostatic interactions. The simulation drives the system toward the minimum of the free energy by evolving the polarization field along the steepest descent of this energy landscape. This gradient flow approach is governed by $\gamma \cdot \frac{\mathrm{d}\hat{\mathrm{n}}}{\mathrm{dt}} = -\frac{\delta \mathrm{F}}{\delta \hat{\mathrm{n}}}$, where $\hat{n}$ is the polar director field, $\gamma$ is a viscosity-like coefficient that sets the rate of relaxation, and $\delta F/\delta \hat{n}$ is the variational derivative, which measures how sensitively the total energy responds to a local perturbation in $\hat{n}$. The negative sign ensures the field evolves in the direction of steepest energy descent.

## VII. Geometry, Discretization & Solver

The 2D domain of the simulation is a $5.0\ \mu\mathrm{m} \times 8.7\ \mu\mathrm{m}$ size rectangle representing a region of a 2.0 μm thick cell, meshed on an equilateral triangular lattice of 64 by 128 cells (see Figure S5). Rows of nodes are placed at a pitch of $\mathrm{h_y} = \mathrm{h}\sqrt{3}/2$; even rows carry 65 nodes at $\mathrm{x} = \mathrm{ih}$ and odd rows carry 64 nodes inset by half a lattice spacing, so that every edge of every triangle has exactly the same length h and no direction in the plane is singled out by the mesh. This matters because the response of the polarization to the charge, including any rotation out of the plane, must not be decided by a mesh anisotropy. The number of rows is even, so that the last row matches the first for the periodic constraint. The polarization and the potential are both discretized with piecewise linear (P1) Lagrange elements on a mixed space.

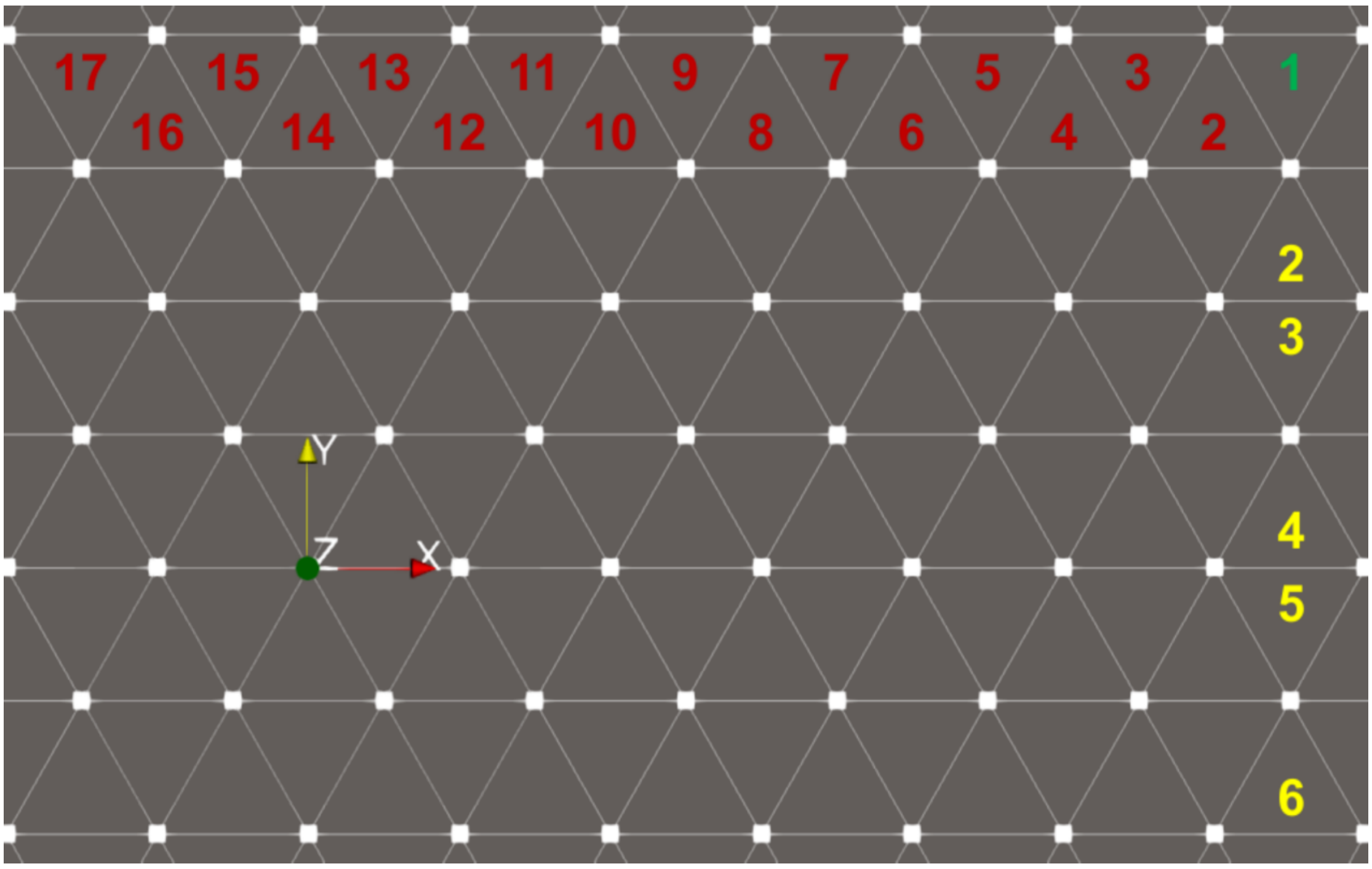

*Figure S5: Sketch of 2D domain of the simulation.*

The gradient flow is advanced by backward Euler with an adaptive step size. At each step the polarization and the potential are solved together as one coupled nonlinear system: the polarization carries a time derivative and the potential does not, so V does not evolve in time; it is recomputed from the Poisson equation at every step. This is the correct separation of scales, since dielectric relaxation in the material is many orders of magnitude faster than director reorientation. The step size starts at 0.1 $ms$ and grows to a maximum of 1 $ms$. A step that fails to converge is rejected, the state is restored, and the step is halved; the step is doubled again only after three consecutive clean solves, since each failed growth attempt costs a full rejected Newton solve. A run aborts if the step falls below 1 ns, which would indicate that the solver, and not the physics, has become the limiting factor.

At each time step the discretized system is nonlinear and cannot be solved by direct matrix inversion. For this reason, Newton's method is employed. Beginning from the previous solution as an initial estimate, the solver iteratively refines the state until the residual, which measures the degree to which the governing equations are not yet satisfied, falls below a prescribed tolerance. Convergence is declared on the residual norm relative to its largest value over the first two iterations of the step, with an absolute floor, or on the size of the Newton update relative to the solution norm. Each Newton iteration reduces to a single linear system, which is solved by direct Lower-Upper factorization using the MUMPS library (12,13).

## VIII. Boundary & Initial Conditions

The domain is periodic along $\hat{y}$, the rubbing axis, which is imposed as a multi-point constraint mapping every degree of freedom on $\hat{y} = L_y$ to its partner on $\hat{y} = 0$, applied separately to each component of $\mathbf{P}$ and to V. The $\hat{x}$ edges are left free, that is, insulating, with the normal component of the electric displacement vanishing. No Dirichlet condition is applied anywhere. The polarization is oriented by the soft plate anchoring free energy penalty and by the two virtual top and bottom layers with polarization set to constant saturation value with director pointing to negative y axis, which enter the energy but carry no degrees of freedom, and the potential is generated purely by the bound and free charges. A pure Neumann Poisson equation with only flux conditions determines $V$ only up to an additive constant, and it is the Debye screening term that removes this freedom, so $V$ is determined without fixing it at an arbitrary point.

The choice of initial condition has a significant consequence for gradient descent simulations. Two systems starting from different initial configurations will, in general, converge to different local energy minima, both in simulation and in experiment. Since the present simulation does not model the phase transition dynamics explicitly, the initial condition is the far-field ground state, where a uniform polarization is antiparallel to the rubbing direction, $\mathbf{P} = \mathbf{P}_S$, so that the texture that develops is the response to the free charge alone. The potential is initialized to zero everywhere.

As noted in Section II, the planar subspace is an exact invariant of these dynamics, so a run started exactly planar, can never leave the plane regardless of how small $K_2$ is. The initial condition therefore adds a small random out of plane component, uniform in the interval $\pm 10^{-3}$ times $P_B$ at each node. If a twist wall forms, it is because the functional prefers one. Setting the seed amplitude to zero turns the model into a strictly planar one and is used as a regression test.

## IX. Material Parameters

The material modeled is DIO in its ferroelectric nematic phase. Table S1 collects the values used.

The Frank constants follow the elastic constant measurements of Paul et al. (14) for a reference ferroelectric nematic material, with the splay constant taken from the recent estimate (15); the large ratio $K_1$ / $K_2$ and $K_1$ / $K_3$ is characteristic of the ferroelectric nematic phase. The dielectric constants and the anisotropic values are reported for DIO (23). We measured the polarization at 65 °C to be 0.038 $C/m^2$ in accordance with Ref (22).

| Quantity | Symbol | Value | Units |
|---|---|---|---|
| Splay Frank constant | $K_1$ | 400.0 | pN |
| Twist Frank constant | $K_2$ | 1.0 | pN |
| Bend Frank constant | $K_3$ | 13.0 | pN |
| Spontaneous Polarization | $P_B$ | 0.038 | $C/m^2$ |
| Parallel Permittivity | $\epsilon_{\parallel}$ | 340 | - |
| Perpendicular Permittivity | $\epsilon_{\perp}$ | 215 | - |
| Bulk Debye length | $\lambda_d$ | 50 | nm |
| Flexoelectric Coefficient | $\gamma_f$ | $-0.004$ | V |
| Polarization Grad. Modulus | $K_h$ | 50 | pN |
| Rotational viscosity | $\gamma$ | 0.05 | Pa s |
| Surface anchoring strength | $W_{Surface}$ | $5 \times 10^{-6}$ | $J\,m^{-2}$ |
| Cell thickness | $L_z$ | 2.0 | μm |
| Domain size | $L_x, L_y$ | 5.00, 8.66 | μm |
| Lattice spacing | h | 78.1 | nm |
| Virtual layer distance | d | 1.0 | μm |
| Lattice | $n_x, n_y$ | 64, 128 | - |
| Free charge | $Q_f$ | $\pm 6.0 \times 10^{-13}$ | C |
| Charge semi-axes | a, b | 1.50, 3.00 | μm |

*Table S1: Material and simulation parameters used.*

Notes: (i) Robustness of the simulation has been checked $\gamma_f$, $K_h$ and $Q_f$ in a reasonable range. (ii) The Debye length $\lambda_d = 50\ nm$ with $\varepsilon_{iso} = 257$ corresponds to about $0.14\ mM$ of each ion species (about $50\ ppm$ molar for DIO, if fully dissociated). For comparison, $0.016\ wt\%$ EMIM-TFSI is about $0.5\ mM$. Accordingly, the value assumes that the nominally pure DIO has about a quarter of the ionic content of the lightly doped sample. (iii) Rest of the values are reasonable values based on prior measurements.

## X. Charge Dynamics

The energy depends on the position of the charge packet only through the term $\rho_f V$, so once the polarization and the potential have been relaxed, the derivative of the energy with respect to the packet position is available exactly by the Hellmann-Feynman argument (24, 25). We define $I$ as integral of $S$ over the sheet, $I_y$ as integral of $dS/dy_c$ over the sheet, $J_0$ as integral of $SV$ over the sheet and $J_y$ as integral of $(dS/dy_c)V$ over the sheet, where $S = (1 - u^2)^2$ is the dimensionless shape of the ellipse. Equating them to 1 at the center $(x_c, y_c)$ and falling smoothly to 0 at its edge, we obtain the generalized force conjugate to the position as $F_{y_c} = -dE/dy_c = -L_z A(J_y - J_0 \frac{I_y}{I})$.

The second term inside the bracket is the contribution from renormalizing the amplitude A against the assembled shape integral. For the profile of $\mathrm{I} = \pi\mathrm{ab}/3$ independent of position for an interior packet, so $\mathrm{I_y}$ vanishes analytically and that term is analytically zero. It is retained nonetheless, because on a mesh $\mathrm{I_y}$ is only zero to quadrature error, and retaining it makes $\mathrm{F_{y_c}}$ the derivative of the discrete energy, which is the energy that the discrete gradient flow decreases.

Given the force, the packet takes one overdamped step per relaxed time step (Stokes' Law),

$$\mathrm{Drag_t}\,\frac{\mathrm{dy_c}}{\mathrm{dt}} = \mathrm{F_{y_c}}, \text{ where } \mathrm{Drag_t} = 6\pi\gamma \mathrm{r_{drag}} \tag{S6}$$

with $\mathrm{r_{drag}}$ the geometric mean of the semi-axes, that is, the radius of a disc of the same area. Relaxation and transport therefore interleave one to one at the same time step. Within a Newton solve the source is frozen, so the packet contributes nothing to the Jacobian. The displacement per step is capped at half a lattice spacing so that a transient spike in the force cannot jump across the domain in a single step.

## XII. Supporting Videos

**Video 1**: Real time temperature dependence of POM textures between crossed polarizers along and perpendicular to the rubbing direction showing the formation and movement of solitons upon cooling from the antiferroelectric phase of a 2.0 μm thick pure DIO sample. The width of the texture shown is 200 μm.

**Video 2**: Real time temperature dependence of POM textures between crossed polarizers along and perpendicular to the rubbing direction showing the formation and movement of solitons upon cooling from the antiferroelectric phase of a 2.0 μm thick DIO sample doped with 0.016% EMIN-TFSI. The width of the texture shown is 200 μm.

**Video 3**: Simulated POM textures between crossed polarizers along and perpendicular to the rubbing direction showing the propagation of solitons screening positive (left) and negative (right) ions in a 2.0 μm thick pure DIO sample.